\documentclass[fleqn,usenatbib]{mnras}
\usepackage{newtxtext,newtxmath}
\usepackage{xcolor}

\usepackage[T1]{fontenc}
\usepackage{pdflscape}

\DeclareRobustCommand{\VAN}[3]{#2}
\let\VANthebibliography\thebibliography
\def\thebibliography{\DeclareRobustCommand{\VAN}[3]{##3}\VANthebibliography}

\usepackage{graphicx}	
\usepackage{amsmath}	
\usepackage{amssymb}	

\title[Low Frequency QPO in MAXI J1820+070]{Low Frequency Quasi-periodic Oscillation in MAXI J1820+070: Revealing distinct Compton and Reflection Contributions}

\author[Gao et al.]{
Chenxu Gao$^{1,2}$,
Zhen Yan$^{1}$\thanks{zyan@shao.ac.cn for correspondence on spectral analysis},
Wenfei Yu$^{1}$\thanks{wenfei@shao.ac.cn for correspondence on timing analysis}%
\\
$^{1}$Shanghai Astronomical Observatory, Chinese Academy of Sciences, 80 Nandan Road, Shanghai 200030, China\\
$^{2}$University of Chinese Academy of Sciences, 19A Yuquan Road, Beijing 100049, China}

\date{Accepted XXX. Received YYY; in original form ZZZ}

\pubyear{2022}

\begin{document}
\label{firstpage}
\pagerange{\pageref{firstpage}--\pageref{lastpage}}
\maketitle

\begin{abstract}
X-ray low frequency quasi-periodic oscillations (LFQPOs) of black hole X-ray binaries, especially those type-C LFQPOs, are representative timing signal of black hole low/hard state and intermediate state, which has been suspected as to originate due to Lense-Thirring precession of the accretion flow. Here we report an analysis of one of the \emph{Insight}-HXMT observations of the black hole transient MAXI J1820$+$070 taken near the flux peak of its hard spectral state during which strong type-C LFQPOs were detected in all three instruments up to photon energies above 150 keV. We obtained and analyzed the short-timescale X-ray spectra corresponding to high- and low-intensity phases of the observed LFQPO waveform with a spectral model composed of Comptonization and disk reflection components. We found that the normalization of the spectral model is the primary parameter that varied between the low and high-intensity phases. The variation in the LFQPO flux at the hard X-ray band ($\gtrsim 100$ keV) is from the Compton component alone, while the energy dependent variation in the LFQPO flux at lower energies ($\lesssim$ 30 keV) is mainly caused by the reflection component with a large reflection fraction in response to the incident Compton component. The observed X-ray LFQPOs thus should be understood as manifesting the original timing signals or beats in the hard Compton component, which gives rise to additional variability in softer energies due to disk reflection. 
\end{abstract}

\begin{keywords}
accretion, accretion discs -- stars: black holes -- X-rays: binaries -- X-rays: individual: MAXI J1820+070
\end{keywords}



\section{Introduction}

Most of black hole X-ray binaries (BH XRBs) in our galaxy are transient sources. They occasionally experienced an outburst and spent most of their time in quiescence. During an outburst, the X-ray spectral and timing properties usually changes dramatically on a timescale of days to months 
\citep{miyamoto_large_1995,remillard_x-ray_2006,done_modelling_2007,belloni_transient_2016-1}. So different spectral states are classified based on the X-ray spectral and timing properties. The two main spectral states are hard and soft states.
The accretion flow in the soft state is thought to be the geometrically thin disk \citep{shakura_black_1973}, which is characterised by the multi-temperature black body X-ray spectrum and very low level short term X-ray variability. The X-ray spectra of the hard state are dominated by the inverse Comptonization scattering of the hot corona. A part of the hard X-ray illuminates the accretion disk and then produces a reflection component, which is characterised by the fluorescent iron K$\alpha$ line and the reflection hump at $\sim$ 30 keV \citep[e.g.][, for a review]{fabian_x-ray_2010}. During the hard state, the X-ray variability amplitude is very high,  and low-frequency quasi-periodic oscillations (LFQPOs) are usually detected. 

One of the most common type low-frequency QPO for BH XRBs is called type-C QPO, the central frequency of which may evolve in the tens of mHz to 30 Hz range, depending on its spectral state\citep[][]{psaltis_correlations_1999,casella_study_2004,van_der_klis_rapid_2006}. The type-C QPO generally appears as very strong and narrow peaks in the power spectrum. The origin of type-C QPO is still under debate, and a popular model is the Lense-Thirring precession of the hot flow\citep[see reviews by][]{ingram_review_2019}. 

Besides conventional timing and spectral analysis of the QPO properties, phase-resolved timing and spectral analysis can provide timing and spectral information of the QPO at different phases, revealing the origin of QPOs \citep{yu_dependence_2001,yu_kilohertz_2002}. Specifically for the LFQPOs in black hole binaries,\citet{miller_evidence_2004} detected variation of the Fe K$\alpha$ line with QPO phase which suggests that the QPOs originate in the inner disk. \citet{ingram_phase-resolved_2015} observed modulation of the Fe K$\alpha$ line width and reflection component by reconstructing the GRS 1915+105 energy spectrum at different phases, providing evidence for a geometric origin of the X-ray QPO. Later on, more comprehensive analysis of the LFQPOs has been performed with the observations of H1743-322 \citep{ingram_quasi-periodic_2016,ingram_tomographic_2017}. Both the iron line energy and the reflection fraction are found modulated with the QPO phase. This is taken as strong evidence supporting the Lense-Thirring precession model.In recent analysis about QPO phase-resolved spectroscopy, \citet{nathan_phase-resolved_2022} detected a significant modulation of the reflection fraction in GRS 1915+105, which indicates that the inner accretion disk geometry changes with QPO phase.
 
MAXI J1820+070 is one of the brightest transient black hole X-ray binaries, which is discovered in X-ray on 2018 March 11 by Monitor of All-sky X-ray Image \citep[MAXI; ][]{kawamuro_maxigsc_2018,tucker_asassn-18ey_2018}. The X-ray intensity of the source increased rapidly and reached a peak of 2 Crab at 2-20keV around March 20 \citep[][]{shidatsu_x-ray_2018}, and then the source stayed in the hard state for roughly three months with slowly decreasing luminosity \citep{shidatsu_x-ray_2019}. This unusually long and bright hard state makes it an excellent target to study the rich soft and hard X-ray spectral and timing phenomena \citep[e.g. ][]{kara_corona_2019,ma_discovery_2020,stiele_timing_2020,you_insight-hxmt_2021}. As optical LFQPOs are observed at frequencies nearly identical to that seen in the X-ray band \citep{yu_detection_2018,yu_further_2018}, multi-wavelength timing observations of the LFQPOs will provide critical clues to the origin of the LFQPOs as well \citep{mao_optical_2022}. 

\emph{Insight}-Hard X-ray Modulation Telescope (HXMT) consists of three payloads: high energy X-ray telescope (HE, 20--250 keV), medium energy X-ray telescope (ME, 5--30 keV), and low energy X-ray telescope (LE, 1--15 keV)\citep{zhang_overview_2020}. The technical details of \emph{Insight}-HXMT can be seen in data reduction guide\footnote{http://hxmten.ihep.ac.cn/SoftDoc/501.jhtml}. \emph{Insight}-HXMT has performed high cadence monitoring observations for this outburst. The broad X-ray energy band and the high count rate during the hard state observations provide us an opportunity to explore the energy spectrum evolution on short timescale. In this work, we use the \emph{Insight}-HXMT data to explore the broadband X-ray spectral variation within the LFQPO timescale.

\section{Data Analysis and Results}

\subsection{Data reduction}
We used the observation taken on March 27, 2018 (ObsId P0114661006), when the flux is near its maximum of the hard state \citep[][]{shidatsu_x-ray_2019}. The exposure time of this observation is about 3400s.

We used the \emph{Insight}-HXMT data analysis software HXMTDAS v2.05 to extract the data from three payloads. To avoid contamination from nearby sources, we chose the narrow FOVs for all three telescopes. For creating good-time-intervals (GTIs), we used the suggested criteria in the guide. We then generated the energy spectrum and the corresponding background from the screened event, as well as the light curve of 8ms resolution.

\subsection{Spectral analysis }
\label{sec:sed}
We grouped the spectra with minimum of 20 counts per bin. 
The spectral fitting performed by \texttt{XSPEC} version 12.11.1 with Chi statistics. We apply this model \texttt{constant$\times$TBabs$\times$(diskbb+reflkerr)} to fit the spectra with energy bands 2--8keV for LE, 10--30keV for ME, 35--150keV for HE. The constant factor is used to account for differences of flux calibration between different payloads. The {\tt\string TBabs} accounts for interstellar absorption, with $N_H$ fixed at $1.5\times10^{21}cm^{-2}$\citep{uttley_nicer_2018}, where abundances and cross sections of the absorption by the Galactic interstellar medium are set according to \citet{wilms_absorption_2000} and \citet{verner_atomic_1996-1}. The {\tt\string diskbb} is a multiple black-body component. The {\tt\string reflkerr} is a relativistic reflection model which includes a thermal Comptonization continuum as the illuminating source ({\tt\string compps}), which is one of the most accurate models of this process, and a hybrid model of rest-frame reflection \texttt{hreflect} \citep{niedzwiecki_improved_2019}. The ionization parameter of the disk $\log\xi$ is defined as the same way as the {\tt\string xillver} \citep{garcia_x-ray_2010}, $\xi=4\pi F_\mathrm{irr}/{n_e}$, where the $F_\mathrm{irr}$ is the irradiating flux in the 13.6 eV-13.6 keV photon energy range and $n_e$ is fixed as $10^{15}cm^{-3}$, \textbf{and without taking into account the ionization gradient.} The comparison between the \texttt{reflkerr} and the other popular relativistic reflection model \texttt{relxill} is extensively shown in \citet{dzielak_comparison_2019}.

During the spectral fitting, we fixed the black hole spin at 0.13 \citep{zhao_estimating_2021} and the inclination angle at $63^\circ$ \citep{atri_radio_2020}. The disk temperature is set as the seed photon temperature for Comptonization. The normalization of the {\tt diskbb} component $N_\mathrm{disk}$ is expressed as $(R_{in}/D_{10})^2\cos\theta$, where $R_{in}$ is the inner disk radius in km, $D_{10}$ is the distance to the source in units of 10 kpc and $\theta$ is inclination angle. We obtained $N_\mathrm{disk}\simeq816(R_\mathrm{in}/R_\mathrm{g})^2$ by taking the distance as 2.96 kpc \citep{atri_radio_2020}, the black hole mass as $8.48M_\odot$ \citep{torres_binary_2020}, and the inclination angle as $63^\circ$ \citep{atri_radio_2020}. Then  $N_\mathrm{disk}$ is tied to the parameter $R_\mathrm{in}$ (in unit of $R_\mathrm{g}$) in {\tt reflkerr} as $816R_\mathrm{in}^2$. The best-fitting parameters are shown in the first row of \autoref{tab:fit_result}. The reduced chi-square value indicates that our model is adequate to describe the data. We also noticed that some spectral fittings need an additional unchanged non-relativistic reflection component during the hard state \citep{buisson_maxi_2019,you_insight-hxmt_2021}. However, whether including the non-relativistic reflection component does not affect the estimation of the relativistic reflection component and the Comptonization component \citep{you_insight-hxmt_2021}.

\begin{figure*}
\includegraphics[width=\linewidth]{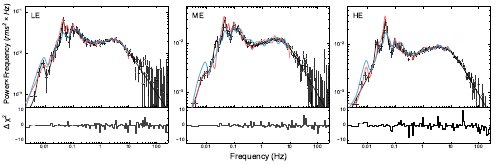}
\caption{Upper: PDSs of LE, ME and HE light curves and faked light curves. The PDS of the original X-ray light curve is shown as black dots and the best-fitting model is shown as black solid line. The PDSs of identified flares (occupy $\sim$ 40\% of the entire exposure) is shown in red, and the PDS of non-flare intervals is shown in blue (for clarity, the uncertainties and the PDS above 10Hz are not plotted). By comparing the PDSs corresponding to the flare and non-flare intervals, we show that the power of the LFQPO is well accounted for by those identified flares. Lower: The contribution to the chi-square of each bin.}
\label{fig:power}
\end{figure*}

\subsection{Timing analysis}
\label{sec:timming}
\begin{figure*}
\includegraphics[width=\linewidth]{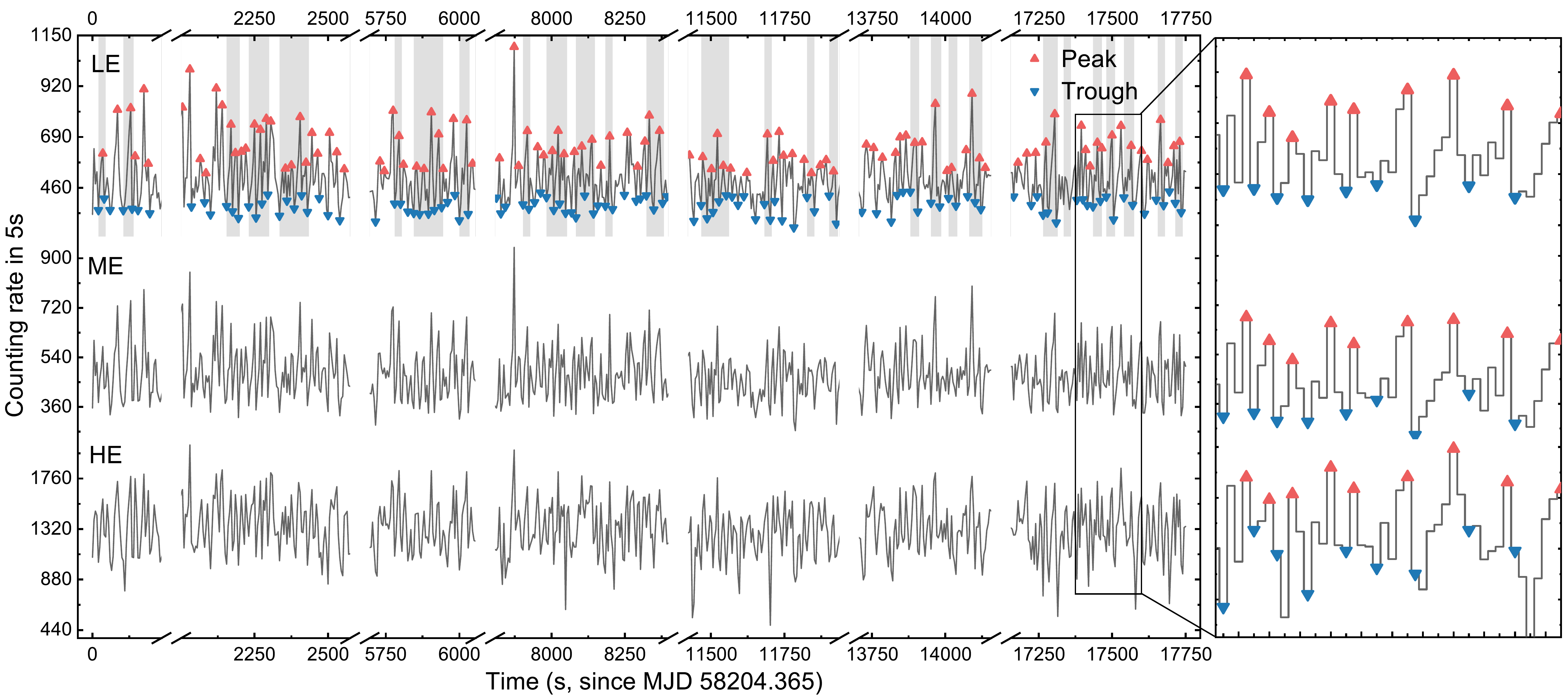}
\caption{The background-subtracted light curves obtained in the observation {\tt\string P011466100601} with a time resolution of 5 seconds. 
LE, ME and HE light curves are shown in the left panel from top to bottom, respectively. The triangle markers in the LE light curves indicate peak (red) and trough (blue) intervals we identified for the LFQPOs, respectively. \textbf{The shaded region in LE light curve indicates the interval of identified flares, the PDS of which shows that the QPO mostly account for those identified flares (see \autoref{fig:power}).} The right panel is a zoom-in on a segment of the light curve.}
\label{fig:0601counts}
\end{figure*}

We generated the power density spectra(PDS) of different energy bands by using \texttt{powspec} with frequency resolution $d\nu = 1/512 $ Hz and Nyquist frequency $\nu_{N} = 256$Hz. We then calculated the rms normalized power by dividing the mean count rate for different energy bands respectively\citep[][and references therein]{van_der_klis_rapid_1995} and subtracted the white noise. We rebined the PDS with a geometric index of 1.08 and fitted the broadband noise with three Lorentzians and the QPO and its harmonic frequencies with three narrow Lorentzians in XSPEC v12.11.1. The best-fitted reduced $\chi^2$ are all less than 1.0. The PDS data and the best-fitting model are plotted in \autoref{fig:power}. The LFQPO around 0.044 Hz is clearly detected as indicated by the curves in black in different energy bands. We show the best-fitting LFQPO results in \autoref{tab:lfqpo}. The central frequency of the LFQPO is constant with the same in different energy bands. The RMS of the LFQPO however decreases with photon energy in the energy range above $\sim$10 keV, which is different from other black hole transients \citep[e.g.][]{huang_insight-hxmt_2018}.

\autoref{fig:0601counts} shows the light curves of three instruments with time resolution of 5 seconds. There are visible flares on the timescales similar to that of the detected LFQPO. In order to quantitatively identify the peak (high intensity) and trough (low intensity) phases of the flares corresponding to the LFQPO waveform, we used the local maxima/minima method to select the extreme from the beginning to the end of the light curve. The method takes a fixed-width window and moves in order from the light curve without overlapping, picking the extreme values of the curve in the window with each move. Since the central frequency of the detected QPO is 0.044 Hz with an FWHM is 0.013 Hz (roughly corresponding to a timescale 20-30 seconds), we then set the window width as 25 s in order to match the QPO cycles. We set the minimum intensity $F_\mathrm{min}$ and the maximum intensity $F_\mathrm{max}$ of the light curve as two reference flux baselines and their intensity difference is $\Delta{F}$. In order to ignore the flares with small fluctuations, we exclude peaks below $F_\mathrm{max}-0.7\Delta{F}$, and troughs above $F_\mathrm{min}+0.18\Delta{F}$. Under the above conditions, we selected 113 peaks and 118 troughs, which 113 pairs of peak and trough phases are adjacent. The selection of the peak phases and the trough phases was only performed in the LE light curve, which is shown in \autoref{fig:0601counts}.

To check whether the peaks and troughs of the selected flares are consistent with the QPO waveform, we produced a power spectrum of a faked light curve consisting of these selected flares. The interval between two neighbor trough is identified as a full flare, which corresponds to the QPO waveform with the time interval less than or equal 30~s. These identified flares covered $\sim$ 40\% of the entire light curve exposure (see the shaded region of \autoref{fig:0601counts}). We then generate two fake light curves based on the identified flare intervals. One is that the count rates during the flare intervals are retained, while the count rates during the non-flare intervals are replaced by the average count rate specific to each interval. The other is that the count rates during the flare intervals are replaced by their corresponding average count rate, while non-flare intervals are retained. The PDSs of the above fake light curves are shown in red and blue solid lines in \autoref{fig:power}. It is obvious that the power spectra of the faked light curve of flare intervals show strong LFQPO. The power at the QPO frequency of 0.044 Hz in the faked PDS of flare intervals is larger than that of the non-flare intervals by a factor of 2.2 for the LE band, 2.0 for the ME band, and 2.0 for the HE band, respectively. At the same time, the band-limited noise components, e.g., the peaked noise in 1--10 Hz, are consistent with the same between that of the flare intervals and that of the non-flare intervals. The observed power of the LFQPO, especially that of the fundamental QPO peak, is mostly accounted for by those identified flares, demonstrating that the peak phase and the trough phase we selected represent those of the QPO waveforms. Thus the differences of the energy spectra corresponding to the peak and the trough phases indeed indicate the spectral variation within the LFQPO cycle.

\begin{table}
\caption{Spectral parameters of the energy spectra corresponding to the LFQPO peak phase, trough phase and the average. All errors indicate 68\% of the uncertainty interval}
\begin{tabular}{lccc}
\hline
Parameter                        & Total                  & Peak                     & Trough                 \\ \hline
$T_\mathrm{in}$ (keV)            &$0.32^{+0.03}_{-0.03}$  &$0.32^{+0.12}_{-0.03}$    &$0.33^{+0.11}_{-0.02}$    \\
$R_\mathrm{in}$ ($R_\mathrm{g}$) &$2.58^{+0.34}_{-0.2}$   &$2.78^{+0.38}_{-0.76}$    &$2.84^{+0.51}_{-0.78}$    \\
$A_\mathrm{Fe}$                  &$3.53^{+0.19}_{-0.35}$  &---                       &---                       \\
$\tau$                           &$4.43^{+0.12}_{-0.17}$  &$4.39^{+0.21}_{-0.1}$     &$4.59^{+0.22}_{-0.15}$   \\
$kT\mathrm{e}$ (keV)             &$44.36^{+1.59}_{-1.08}$ &$45.1^{+1.24}_{-2.18}$    &$43.6^{+0.96}_{-1.62}$  \\
log$\xi$                         &$3.78^{+0.06}_{-0.04}$  &---                       &---                     \\
$\mathcal{R}$                    &$1.18^{+0.07}_{-0.04}$  &$1.28^{+0.1}_{-0.1} $     &$1.05^{+0.1}_{-0.11} $   \\
$norm$                           &$1.65^{+0.09}_{-0.11}$  &$2.03^{+0.14}_{-0.35}$    &$1.0^{+0.09}_{-0.21}$    \\ 
$\chi^2$/dof                         &$1138/1235$             &$1286/1226$              &$1412/1184$               \\ \hline
\label{tab:fit_result}
\end{tabular}
\end{table}

\begin{table}
\caption{Best-fitting results for LFQPO around 0.044 Hz in different energy bands}
\begin{tabular}{lcccccc}
\hline
                    & $\nu_\mathrm{QPO}$ (Hz)       & FWHM                      & RMS                        &$\chi^2$/dof  \\ \hline
2-8keV(LE)          &$0.044^{+0.001}_{-0.002}$ &$0.013^{+0.005}_{-0.005}$  &$0.156^{+0.026}_{-0.025}$   &146/136\\
10-30keV(ME)        &$0.043^{+0.002}_{-0.001}$ &$0.014^{+0.005}_{-0.004}$  &$0.106^{+0.023}_{-0.018}$   &132/136\\
35-150keV(HE)       &$0.044^{+0.001}_{-0.001}$ &$0.01 ^{+0.004}_{-0.003}$  &$0.093^{+0.007}_{-0.013}$   &159/136\\
100-150keV(HE)      &$0.044^{+0.002}_{-0.001}$ &$0.009^{+0.003}_{-0.002}$  &$0.08^{+0.007}_{-0.007}$    &138/136\\ 
150-200keV(HE)      &$0.044^{+0.004}_{-0.003}$ &$0.007^{+0.006}_{-0.005}$  &$0.06^{+0.02}_{-0.01}$  &140/136\\
\hline
\label{tab:lfqpo}
\end{tabular}
\end{table}

\subsection{Short-timescale energy spectra of the LFQPO: the peak phase vs. the trough phase}

Then we performed spectral fits to all the energy spectra on 5 second time scales which were extracted in the time intervals corresponding to the peak phase and the trough phase, respectively. For each energy spectrum, we performed Markov Chain Monte Carlo sampling in XSPEC, with the sampling algorithm from the \texttt{emcee} software package \citep{foreman-mackey_emcee_2013}. We fixed the following three parameters: the constant value, the iron abundance $A_\mathrm{Fe}$, and the ionization parameter $\log{\xi}$ to the same as the average energy spectrum. We set 20 walkers with the prior parameters from the average energy spectrum fitting result in \autoref{sec:sed}. We used a Gaussian prior with a center as the best-fitting value from the parameters of the averaged energy spectrum and a width as 20\%. We set 5,000 walkers and 5,000 burn-in steps. The median value of the posterior samples is taken as the best-fitting value for each free parameter. The 68\% interval of reduced chi-square extracted from the median value of posterior chi-squares is 0.93 to 1.1, which indicates that all the 5s energy spectra are well-fitted by the model.

\begin{figure*}
\includegraphics[width=\linewidth]{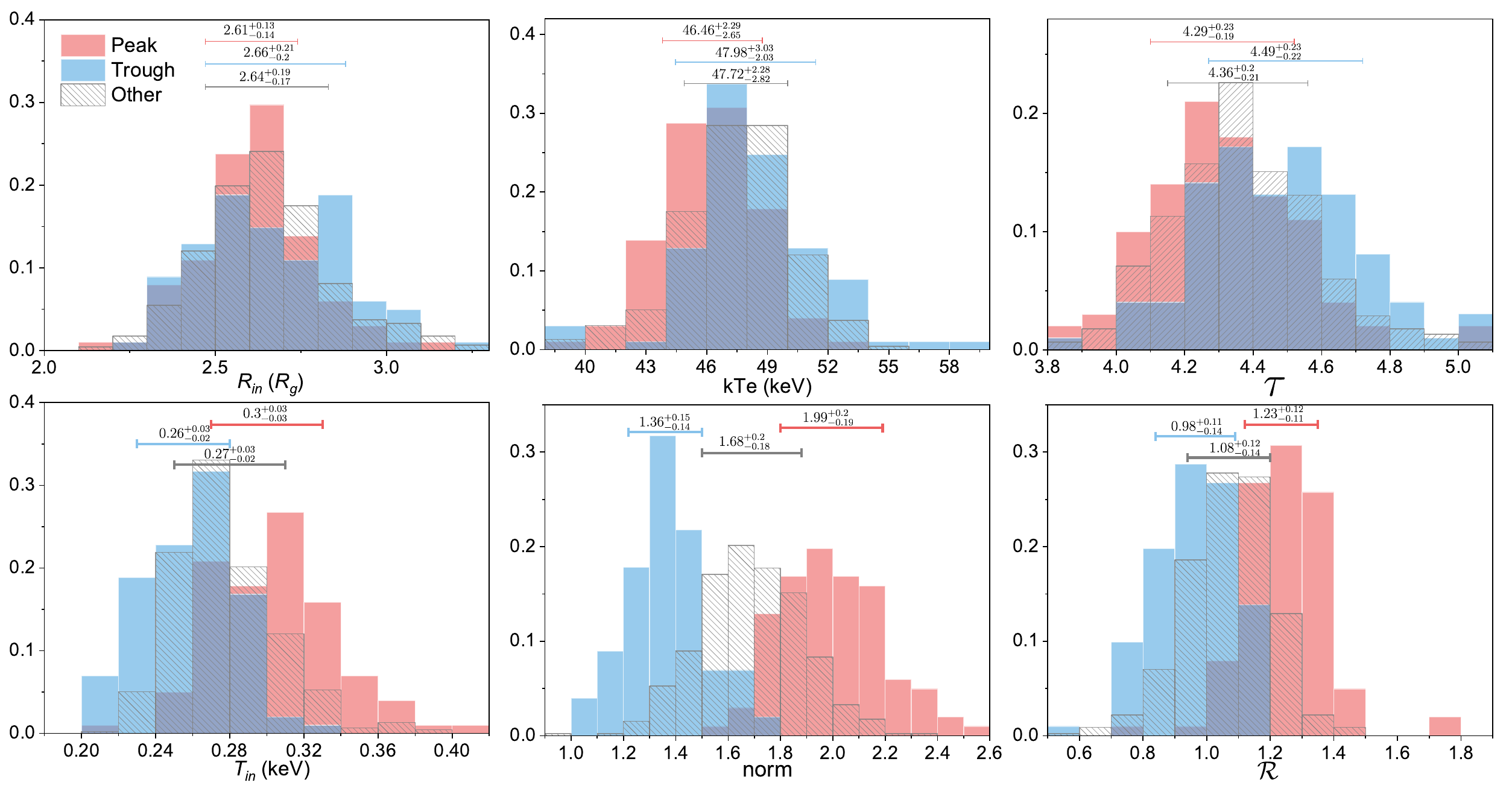}
\caption{The histogram of the best-fitting parameters of all the X-ray spectra extracted on 5 second time scale. In each graph, the histograms in red represent those spectra during the peak phase, while the histogram in blue represents those spectra during the trough phases, respectively. The histograms of the spectral parameters in neither the peak phase nor the trough phase are shown in shadow. The 68\% interval for each parameter in each histogram is marked by a horizontal baa r.
\label{fig:histogram}}
\end{figure*}

We then obtained the best-fitting parameters for all the X-ray spectra extracted on the 5 second time scale. The histogram of the best-fitting spectral parameters in the peak, trough and other phases are plotted(\autoref{fig:histogram}). Then the mean value and 68\% interval (16\%-84\%) for each parameter of each phase are also calculated. We can see that the inner disk radius $R_\mathrm{in}$, the electron temperature $kT_\mathrm{e}$ and the optical depth $\tau$ show almost identical distribution between peak and trough. The other three parameters are larger in the peak phase than in the trough phase. Especially, the 68\% intervals of the normalization $norm$ and the reflection fraction $\mathcal{R}$ show no overlap, indicating a significant difference between these two parameters between peak and trough phases. According to the results, we obtain that the peak and trough phase short timescale spectral parameters $T_{in}$, $\mathcal{R}$ and $norm$ are different at least 50\%, 76\% and 96\% significance level. So at least the spectral parameter $norm$ significantly changes with the QPO phases . We also generated the respective PDSs for all best-fitting parameters of the short timescale spectra, but only the PDS of the $norm$ shows an apparent QPO feature at the same frequency.

The most significant difference between the energy spectra corresponding to the peak and trough phases is the normalization (see \autoref{fig:histogram}). The normalization determines the photon flux of both Compton and reflection components by the same factor. The reflection fraction $\mathcal{R}$ also varies between peak and trough phases. The photon flux of the reflection component is determined by $norm\times \mathcal{R}$ \citep[][]{niedzwiecki_improved_2019}. So the variation of $\mathcal{R}$ makes the flux variability of the reflection component different from that of the Compton component. The normalization and reflection fraction are also found to significantly vary with the QPO phase in another BHXRB H1743$-$322, in which the normalization is also the parameter that bears the most significant change  \citep{ingram_tomographic_2017}. There is also difference in the inner disk temperature $T_\mathrm{in}$ between that of the peak and of the trough. Since the disk component was very weak in the observation and the inner disk temperature is tied to the seed photon temperature for Compotonization in the model we applied, we can not distinguish whether the variation in $T_\mathrm{in}$ was due to intrinsic disk variability or potential variation in the Compton component.

\subsection{Averaged energy spectra corresponding to the LFQPO peak and trough phases}
\begin{figure*}
\includegraphics[width=\linewidth]{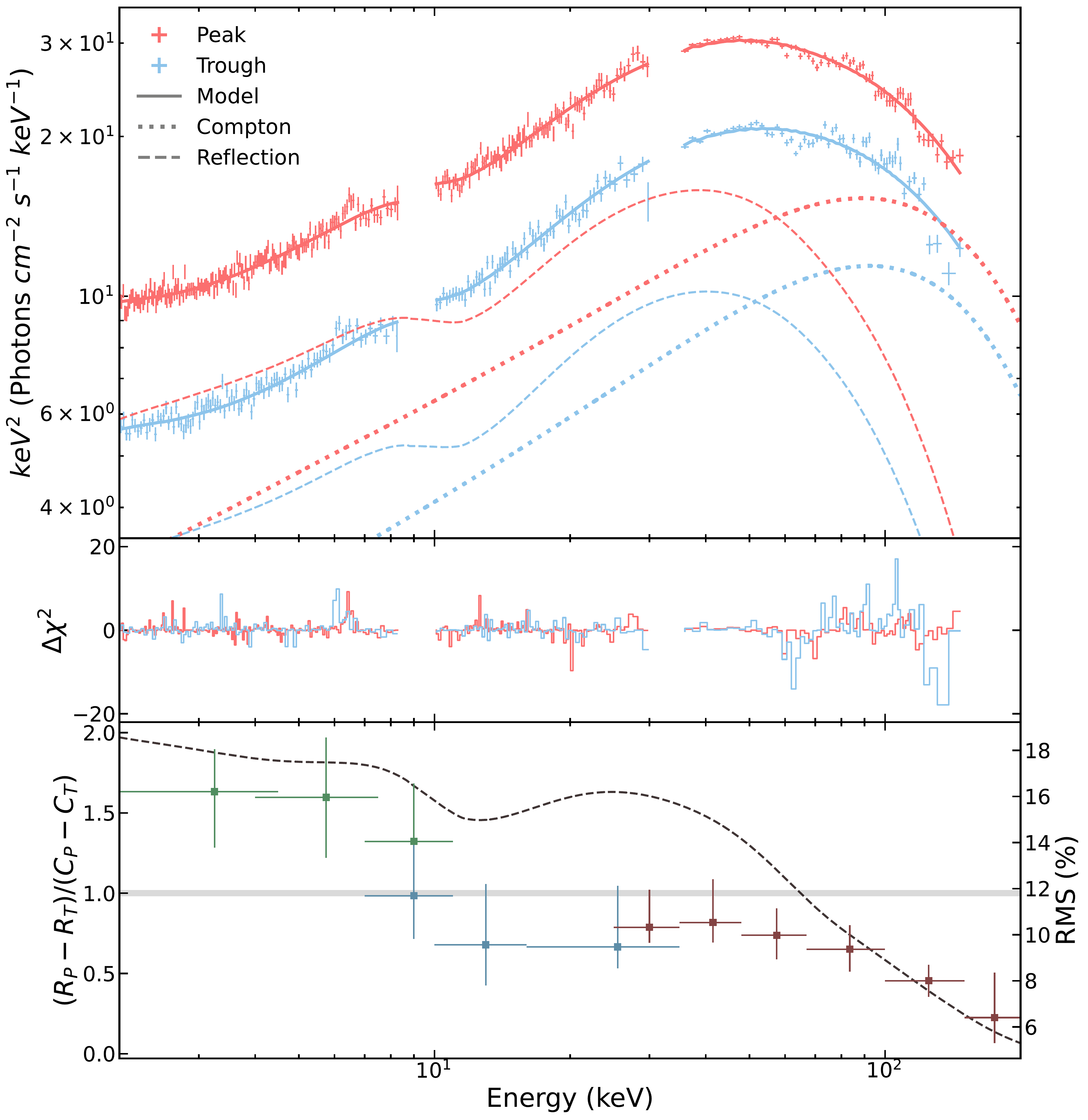}
\caption{Upper: The unfolded energy spectra corresponding to the peak phase (in red) and the trough phase (in blue) and the corresponding spectral components of the best-fitting models. The solid line shows the total spectrum. The disk component is not shown as it is very weak and negligible. The Compton and reflection components are represented by a dotted line and a dashed line, respectively. Middle: The contributions to the chi-square for the peak and trough spectra. Lower: The ratio between the flux changes between the peak phase and the trough phase in the Compton component and the reflection component (shown as the dashed line), and the QPO fractional RMS vs. energy up to 200 keV (marked in solid square at the lower panel). Here $R_\mathrm{P}$-$R_\mathrm{T}$ is the photon flux change of reflection component, and $C_\mathrm{P}$-$C_\mathrm{T}$ is the photon flux change of Compton component.
\label{fig:component}}
\end{figure*}

In order to compare the energy spectra corresponding to the peak and trough phases in detail with enough statistics, we stacked all the 5-second energy spectra corresponding to either phase. Corresponding response files were generated using the tool \texttt{ADDSPEC} and the background file was generated using the tool \texttt{MATHPHA}, respectively. We also used the same spectral model that was used to fit the average energy spectrum to fit the stacked energy spectra. We fixed the constant value, the iron abundance $A_\mathrm{Fe}$ and the ionization parameter $\log{\xi}$ in the model at the same values obtained in the spectral fit of the average energy spectrum. The best-fitting spectral parameters corresponding to the peak and the trough phases are listed in \autoref{tab:fit_result}. 
The best-fitting spectral parameters corresponding to the peak and trough phases agree with each other, except for the normalization $norm$, the reflection fraction $\mathcal{R}$. The $norm$ and $\mathcal{R}$ of the peak and trough phases are different by at least 97\% and 70\% significance levels, which is consistent with the statistical results of the 5-seconds spectra. 

We plotted the unfolded energy spectra of peak/trough phases in the top panel of \autoref{fig:component}. The best-fitting models of the Compton and reflection components are also added in the \autoref{fig:component}. The spectral shape of the Compton components are almost the same at the peak and the trough phases, since only $norm$ changes and other parameters related to Compton (such as $kT_\mathrm{e}$ and $\tau$) remain nearly constant. On the other hand, the spectral shape of the reflection component at the peak phase is different from that at the trough phase, since the reflection fraction $\mathcal{R}$ is larger. The reflection component dominates over the Compton below $\sim$30 keV, and the Compton component becomes dominated above $\sim$70 keV in both peak and trough phases.   

The fractional photon flux change in the Compton component between peak and trough phases ($C_\mathrm{P}$-$C_\mathrm{T}$) is almost the same for different energies since the normalization is the only parameter that varies. The photon flux variation in the reflection component ($R_\mathrm{P}$-$R_\mathrm{T}$) is roughly two times larger than that of the Compton component below $\sim$10 keV, and becomes less than 1/2 in the energy range above $\sim$100 keV. We then plotted the ratio between the photon flux variation of the reflection and the Compton component between the peak phase and the trough phase, in order to show relative contributions from the Compton and the reflection component to the X-ray variability with photon energy. The QPO fractional RMS below $\sim$10 keV is also larger than that above $\sim$10 keV (see \autoref{fig:component}), which suggests that strong variation of the QPO flux at lower energy band is due to the reflection component. Since the photon flux variation of the reflection component is determined by the variation of $norm\times \mathcal{R}$, the dominated role of the reflection component below $\sim$10 keV on the QPO flux variation can be understood as the parameter $\mathcal{R}$ of this observation is large. 

A remarkable result is shown in \autoref{fig:component}. Starting from above $\sim$70 keV, the flux variation between the peak phase and the trough phase is becoming primarily from the contribution of the {\it direct} Compton component. Beyond 100 keV or so, the contribution of the reflection component is negligible. This implies that detection of the LFQPOs at energy above $\sim$100 keV simply means that the Compton component itself produces the LFQPO quasi-periodicity, although the reflection component dominates the LFQPO peak-trough variation below $\sim$30 keV. 

\section{Discussion}
We have investigated the short-timescale X-ray spectra corresponding to the high and low intensity phases of the LFQPO (defined as peak and trough) in one $Insight$-HXMT observation of MAXI J1820+070. 
The greatest change among spectral parameters between the two LFQPO phases is the normalization $norm$. The difference in the normalization represents the flux variation between the two phases in the primary Compton emission. In the energy range above $\sim$100 keV, the Compton component overwhelmingly dominates over the reflection component, and thus the flux variation above $\sim$100 between the peak and the trough phases is contributed by the Compton component alone (see \autoref{fig:component}). Therefore, the underlying beats that generate the LFQPOs should originate from the Compton emission. Under the framework of conventional Comptonization models, the Compton component is thought to originate from a hot corona. Our spectral analysis shows that the corona is not oscillating coronal temperature or electron density, since in our spectral fits, the temperature $kT_{e}$ and $\tau$ of the Compton component does not show apparent changes between those of the peak and the trough phases. Thus the LFQPO corresponds to a modulation in the photon flux of the Compton component, either through modulation of the seed photons or line-of-sight covering factor. 

The difference in the reflection fraction $\mathcal{R}$ between the peak and the trough phases demonstrates that the reflection fraction acts like an amplification factor of the variation in the primary incident Compton flux since the photon flux variation in the reflection component is determined by $norm\times\mathcal{R}$. The values of the reflection fraction in the peak and the trough phases are both very large (larger than unity) but different, which causes the flux change between the peak and the trough phases at lower energies (below $\sim$30 keV, see \autoref{fig:component}) is dominated by the reflection component. The effect of the reflection fraction is also confirmed by that the fractional rms of the LFQPOs at below $\sim$10 keV is higher than that above 10 keV. In conclusion, the modulation in the Compton photons produces the underlying, original beats of the LFQPO signal, while the LFQPO amplitude at lower energy bands is primarily contributed by the reprocessed reflection component, which is affected and amplified by the reflection fraction. 

Hot accretion flow model has been the popular model to account for the Compton emission in the hard state of BHXRB \citep{done_modelling_2007}. In this model, the accretion geometry is composed of a hot accretion flow inside a truncated disk \citep{esin_advection-dominated_1997}. The precession of the hot accretion flow can modulate the Compton emission through self-occultation, projected area and relativistic effects, which produce the observed LFQPO \citep{ingram_low-frequency_2009}. Because there is a misalignment between the hot accretion flow and the accretion disk, the reflection fraction changes with the hot accretion flow precession. The variation of the reflection fraction with LFQPO phases strongly depends on the inclination angle and the truncated disk radius \citep{you_x-ray_2020}.  

In an alternative accretion geometry, the so-called lamppost geometry, a point corona lies above the black hole which illuminates the accretion disk. If the precession of the disk is the proposed mechanism to produce the LFQPO under this accretion geometry \citep[][]{schnittman_precessing_2006}, the modulation in the Compton component at high energy band (say, e.g., above 100 keV) is not expected unless the corona is partially obscured by the precessing disk.

Besides, oscillation of the corona has also been proposed to explain the observed LFQPOs \citep{cabanac_variability_2010}. However, the variations of the corona properties such as the electron temperature predicted by this model was not observed. In addition, an increasing trend of the RMS variability with energy relation predicted by this model is not seen either. Our results rule out the oscillation of the corona as the cause of the LFQPOs. 

We have also noticed that the Two-Comptonent Advection Flow (TCAF) is used to explain the X-ray spectral and timing properties of black hole binaries in the hard state. The post-shock region in this model acts as a Compton cloud, and the oscillation of the Compton cloud causes the LFQPO\citep{chakrabarti_evolution_2008}. The Keplerian disk is situated outside the shock location and should be very far away from the central black hole (hundreds of $r_\mathrm{g}$) in order to match the observed LFQPO frequency\citep{debnath_implementation_2014}. It would be very difficult to produce a strong disk reflection component and the potential oscillation of the Compton cloud further out in such a scenario. 
 
Jet can also act like a corona which produces Compton photons \citep{markoff_going_2005}. MHD simulations have shown that a precessing jet can be formed and precession of the jet might be responsible for the LFQPO \citep{liska_formation_2018}. Our results are rather consistent with such an idea that the Compton emission from the jet or jet base illuminates the accretion disk and produces the reflection component \citep[e.g. ][]{dauser_irradiation_2013,kara_relativistic_2016}. Other proposals which attribute the broadband X-ray LFQPOs as due to processing jet alone \citep{ma_discovery_2020,ferreira_are_2022} is not supported by our spectral analysis, as a reflection component with strong effect of reflection fraction is required. 

\section{Conclusion}
By performing LFQPO phase-dependent spectral analysis, we found the most varied parameter between the energy spectra corresponding to the peak phase and the trough phase of the black hole LFQPOs is the normalization $norm$ in the Comptonization model we applied, and the second most varied parameter is the reflection fraction $\mathcal{R}$. Both the Compton and reflection spectral components contribute to the LFQPOs observed in broadband X-rays. In the spectral model applied in our spectral analysis, the difference in the normalization $norm$ represents the variation in the Compton emission component, and the parameter reflection fraction $\mathcal{R}$ serves as an apparent amplification of the incident Compton component into the resulted reflection component, adding additional modulation in the reflection component with the LFQPO phase. Our investigation {implies} that the {original} timing signals or beats that are responsible for the LFQPOs lie in the Compton emission {component}, and the energy-dependent behavior of the LFQPOs at softer energies ($\lesssim$ 30 keV) is the result of the reflection component with a rather large reflection fraction. If the LFQPOs are due to Lense-Thirring precession, the phase information of precession then lies in the Compton component rather than in the reflection component.                                                                                                                                                  

\section*{Acknowledgements}
We would like to thank Andrzej Niedzwiecki and Bei You for the helpful discussion about the reflection model. This work was supported in part by the National Natural Science Foundation of China (grant Nos. U1838203). Z.Y. was supported in part by the National Natural Science Foundation of China (grant Nos. U1938114), the Youth Innovation Promotion Association of CAS (id 2020265) and funds for key programs of the Shanghai Astronomical Observatory.

\section*{Data Availability}
The public data used in this work can be downloaded in the \emph{Insight}-HXMT official website (\url{http://hxmten.ihep.ac.cn/})




\bibliographystyle{mnras}
\bibliography{cxgao_lib} 








\bsp	
\label{lastpage}
\end{document}